\documentclass[aps,prd,notitlepage,nofootinbib,superscriptaddress,amsmath,amssymb,showpacs,longbibliography]{revtex4-2}

\usepackage{graphics}
\usepackage{graphicx}
\usepackage{color}
\usepackage{hyperref}
\usepackage{bm}
\usepackage{calligra}
\usepackage[T1]{fontenc}
\usepackage{amsmath}
\usepackage{amssymb}
\usepackage{amsfonts}
\usepackage{mathrsfs}  
\usepackage{mathtools}
\usepackage{slashed}
\usepackage[normalem]{ulem}
\usepackage{dsfont}

\usepackage[utf8]{inputenc}

\usepackage{lipsum}

\usepackage{textcomp}
\usepackage{color}

\usepackage{lineno}
\DeclareMathOperator{\tr}{tr}

\DeclareMathOperator{\diag}{diag\,}

\definecolor{darkgreen}{rgb}{0.2, 0.3, 0.1}

\begin{document}

\title{Insights on the peak in the speed of sound of ultradense matter}

\author{M.~{Hippert}}
\email{hippert@illinois.edu}
\affiliation{Illinois Center for Advanced Studies of the Universe,\\ Department of Physics, 
University of Illinois at Urbana-Champaign, 1110 W. Green St., Urbana IL 61801-3080, USA}

\author{E.~S.~{Fraga}}
\email{fraga@if.ufrj.br}
\affiliation{Instituto de F\'\i sica, Universidade Federal do Rio de Janeiro,
Caixa Postal 68528, 21941-972, Rio de Janeiro, RJ, Brazil}

\author{J.~{Noronha}}
\email{jn0508@illinois.edu}
\affiliation{Illinois Center for Advanced Studies of the Universe,\\ Department of Physics, 
University of Illinois at Urbana-Champaign, 1110 W. Green St., Urbana IL 61801-3080, USA}

\date{\today}

\begin{abstract} 
In this work we investigate the minimal physical requirements needed for generating a speed of sound that surpasses its asymptotic conformal limit. It is shown that
a peak in the speed of sound of homogeneous matter naturally emerges in the transition from a phase with broken chiral symmetry to one with a gapped Fermi surface. We argue that this could be relevant for understanding the peak in the speed of sound displayed by some of the current models for cold ultradense matter. A minimal model implementation of this mechanism is presented, based on the spontaneous breakdown of an approximate particle-antiparticle symmetry, and  its thermodynamic properties are determined.   
\end{abstract}

\maketitle


\section{Introduction}

Neutron star observations provide a unique window into the physics of cold ultradense matter. 
Recent measurements of masses $\gtrsim 2\,M_\odot$ \cite{Demorest:2010bx,Antoniadis:2013pzd,Cromartie:2019kug}, for instance, 
provide evidence of a stiff equation of state (EoS) in the core of these stars, posing important constraints to theoretical models of dense neutron-star matter \cite{Buballa:2014jta,Kurkela:2014vha,Annala:2017llu,Nandi:2018ami}. In particular, when combined with knowledge from nuclear physics at lower densities, such observations seem to challenge the conjecture of a conformal bound for the speed of sound $c_s\leq 1/\sqrt{3}$ \cite{Bedaque:2014sqa,Alsing:2017bbc,Tews:2018kmu,Ma:2018qkg,Zhang:2019udy,Alford:2015dpa} (we use natural units $\hbar=c=k_B=1$). 
Further support against the bound came from the gravitational-wave event GW170817 \cite{TheLIGOScientific:2017qsa,Abbott:2018exr,Abbott:2018wiz}, associated with a binary neutron-star merger, which provided new and independent constraints on the masses and effective tidal deformability of the inspiraling neutron stars \cite{Tews:2018kmu,Tews:2019cap,Reed:2019ezm,Capano:2019eae,Annala:2019puf,Kanakis-Pegios:2020jnf,Huth:2020ozf,Han:2021kjx}. Since in QCD at asymptotic densities one expects $c_s^2 \to 1/3$, the need for surpassing the 1/3 mark at intermediate densities implies that the speed of sound is not monotonic and that there is at least one maximum, or a peak, in the speed of sound of cold ultradense matter as a function of density. 

More recently, the observation of gravitational waves from the merger of a black hole with a compact object of mass $2.6 \,M_\odot$ \cite{Abbott:2020khf} has raised speculations on the possibility of neutron stars with masses above $2.5\,M_\odot$, which would pose an even greater challenge to current knowledge on the EoS of cold and dense matter \cite{Drischler:2020fvz,Tews:2020ylw,Kanakis-Pegios:2020kzp,Adam:2020yfv,Tan:2020ics,Fattoyev:2020cws,Nathanail:2021tay,Tsokaros:2020hli,Lim:2020zvx,Dexheimer:2020rlp,Godzieba:2020tjn,Huang:2020cab,Sedrakian:2020kbi}. 
Regardless of whether such speculations prove right or not, the fact is that accumulating observations will inevitably pose increasingly tight constraints on the structure of neutron stars, and guidance is needed for accommodating these into our knowledge of the dense matter equation of state.  
In particular, crucial information is expected from ongoing measurements on the radius of the extremely massive pulsar PSR J0740+6620, by NICER \cite{Guillot:2019vqp}. 
In light of current observational advances \cite{TheLIGOScientific:2017qsa,Abbott:2018exr,Abbott:2018wiz,Abbott:2020khf,Nattila:2015jra,Nattila:2017wtj,Watts:2018iom,Riley:2019yda,Miller:2019cac,Dexheimer:2020zzs}, it seems plausible that future constraints on the masses, radii, and deformabilities of neutron stars will provide extremely compelling evidence in favor of models for the EoS where $c_s$ goes above the conformal limit \cite{Moustakidis:2016sab,Margaritis:2019hfq,Tews:2020ylw,Drischler:2020fvz,Marczenko:2020wlc,Greif:2020pju,Han:2020adu}. If not found in the extremely dense interior of neutron stars, it seems unlikely that violations of the conformal limit might be found elsewhere in the universe.

The discussion of an upper bound to the speed of sound 
has a long history \cite{HARTLE1978201,Zeldovich:1962emp,Bludman:1968zz,zel2014stars}. 
It is well known that causality forbids the existence of perfectly rigid bodies in the relativistic regime, which implies that the speed of sound must be bounded by the speed of light $c_s\leq 1$ \cite{HARTLE1978201,zel2014stars}. However, not even a gas of photons can achieve this limit given that perturbations in a gas of noninteracting massless particles propagate at $c_s^2=1/3$, and the inclusion of nonzero masses and perturbative interactions lead to even lower speeds  \cite{LANDAU197566,zel2014stars,Graf:2015tda}.  In fact, at high temperatures and zero baryon chemical potential holographic \cite{Gubser:2008yx,Cherman:2009tw,Finazzo:2014cna} and lattice QCD calculations \cite{Borsanyi:2013bia,Bazavov:2014pvz} do find that $c_s^2 \leq 1/3$. 
Nonetheless, speeds of sound above the conformal result can be found in QCD at large isospin density \cite{Carignano:2016lxe}, in two-color QCD \cite{Hands:2006ve}, in holographic approaches \cite{Hoyos:2016cob,Ecker:2017fyh,Anabalon:2017eri,Chesler:2019osn}, in resummed perturbation theory \cite{Gorda:2014vga,Fujimoto:2020tjc}, in quarkyonic matter \cite{McLerran:2018hbz,Jeong:2019lhv,Zhao:2020dvu,Margueron:2021dtx,Duarte:2021tsx} and in models of high-density QCD \cite{Blaschke:2007ri,Kojo:2014rca,Leonhardt:2019fua,Baym:2019iky,Roupas:2020nua,Malfatti:2020onm,Ayriyan:2021prr,Fadafa:2019euu,Xia:2019xax,Stone:2019abq,Li:2020dst,Pisarski:2021aoz,Pal:2021qav,Motta:2021xwo,Ma:2021zev}.
At any rate, it is remarkable that questions regarding the maximum speed of sound allowed in nature might be answered, in the near future, by astrophysical observations.

In this work, we examine the minimal physics ingredients required to evade the conformal result and build a simple mechanism that gathers all these ingredients. We argue that such a mechanism naturally emerges in the transition from a phase with broken chiral symmetry to one with broken $U(1)_B$ symmetry, due to the condensation of diquarks or dibaryons, which for simplicity we call here the ``chiral-superfluid'' transition. This mechanism may be relevant for QCD with two colors \cite{Hands:2001ee,Kogut:2001na,Ratti:2004ra,Hands:2006ve,Hands:2010gd,Boz:2019enj,Astrakhantsev:2020tdl}. We also note that a diquark gap is often present in models featuring a maximum in the speed of sound \cite{Kojo:2014rca,Roupas:2020nua,Leonhardt:2019fua,Baym:2019iky,Malfatti:2020onm,Ayriyan:2021prr}. In fact, in Ref.~\cite{Leonhardt:2019fua} in particular, a functional-renormalization-group calculation was carried out and it was found that the inclusion of a diquark gap is tightly connected with the emergence of a maximum in the speed of sound. Alternatively, in Ref.~\cite{Ayriyan:2021prr}, a Bayesian analysis of current observational data was shown to also favor the presence of diquarks at large densities. We believe our arguments may be helpful to shed more light on these results. 

The outline of this paper is as follows. In Sec.~\ref{sec:generalarguments}, starting from elementary observations, we devise a simple  mechanism capable of meeting all the requirements needed to produce a large peak in the speed of sound. 
We present in Sec.~\ref{sec:symmetry} a possible symmetry-breaking pattern  for the chiral-superfluid transition, based on an approximate, continuous particle-antiparticle symmetry. In Sec.~\ref{sec:model} we construct a minimalist mean-field toy-model implementation of this symmetry-breaking pattern and calculate its thermodynamical properties. Results are presented and discussed in Sec.~\ref{sec:results}.  
Finally, Sec.~\ref{sec:conclusions} presents our conclusions and outlook.

\section{Speed of Sound beyond the conformal limit}
\label{sec:generalarguments}

The approximately conformal speed of sound displayed by QCD at asymptotically large densities follows, in a straightforward manner, from dimensional analysis and asymptotic freedom. At very large chemical potentials, $\mu \gg \Lambda_\text{QCD}$, all other energy scales, as well as interactions, are rendered irrelevant and the pressure $p$ must become proportional to $\mu^{d+1}$, where $d$ is the number of spatial dimensions. In this limit, the speed of sound squared becomes $c_s^2=dp/d\epsilon\sim 1/d$ --- or $c_s^2 = 1/3$, for three spatial dimensions \cite{Zeldovich:1962emp}. At lower densities, mass scales become relevant and the pressure becomes $p\sim \mu^{d+1} (\mu/m)^\alpha$ so that $c_s^2 \sim (d + \alpha)^{-1}$, but, since increasing the mass will generally decrease the pressure, $\alpha > 0$ and the speed of sound becomes subconformal: $c_s^2<1/d$. 

There are two simple ways to evade the argument from dimensional analysis: a) decreasing the number of effective dimensions or b) providing an energy scale $\Phi$ such that  $p\sim \mu^{d+1} (\mu/\Phi)^\alpha$ with $\alpha <0$. For instance, a speed of sound $c_s^2\sim 1$ is found for  $p = p_0(\mu) + q^2\, \mu^2\, \Phi^2/2$ if the energy scale $\Phi \gg p_0^{1/2}/\mu$ is approximately independent of $\mu$. That is, as long as the susceptibility $\chi_2 \equiv d^2p/d\mu^2$ is approximately constant. However, such an energy scale must emerge nonperturbatively and, from here on, we will assume that it arises from the condensation of a (composite) bosonic field. This condensate must increase with chemical potential, starting from a threshold value $\mu=\mu_c$ up to a point $\mu \sim \mu_\mathrm{sat}$ where it saturates and becomes approximately constant. At the same time, its leading contribution to the pressure must be $\sim q^2\,\mu^2\,\Phi^2/2$. It follows that this condensate must carry (baryon) charge  $q$ and, to become approximately constant, there must be somehow a mechanism that keeps it from further increasing --- for instance, some type of energy barrier. 

Reference~\cite{Bedaque:2014sqa} discussed the connection between the conformal speed-of-sound limit, $c_s^2\leq 1/d$, and the effective degrees of freedom relevant for the system at a given density. In fact, they argued that the effective number of (fermionic) degrees of freedom $N_{\textrm{eff}}(\mu) \propto n/\mu^d$, where $n$ is the density, must decrease with $\mu$ to produce a speed of sound $c_s^2>1/d$. We note that this naturally occurs in the presence of a saturating baryon-number-carrying condensate: because the condensate stops growing at $\mu \sim \mu_\mathrm{sat}$, the corresponding degrees of freedom 
``freeze'' and, thus, $dN_{\textrm{eff}}/d\mu<0$. 

Of course, a mechanism capable of saturating a baryon-number-charged condensate once it has achieved a sufficiently large value is needed to create a regime where $dN_{\textrm{eff}}/d\mu<0$. This cannot be achieved by conjecturing a simple quartic mean-field potential for $\Phi$ alone. 
In fact, if we consider an effective potential  $\Omega(\Phi,\mu) = \Omega_0(\Phi) -q^2\,\mu^2\,\Phi^2/2$ to describe this physics, with $\partial \Omega/\partial\Phi=0$, then from the total derivative 
\begin{equation}
    \frac{d\;}{d\mu}\frac{\partial \Omega}{\partial \Phi} = -2\,q^2\,\mu\,\Phi+ m_\Phi^2\,\frac{d\Phi}{d\mu} =0\,,
    \label{eq:totalmuderivative}
\end{equation}
one finds
\begin{equation}
    \frac{d\Phi}{d\mu} = 2\,q^2\,\frac{\mu\,\Phi}{m_\Phi^2}\sim 2\,q^2\,v^{\gamma-4}\,\mu\,\Phi^{3-\gamma}\,,
\end{equation}
where $m_\Phi^2\equiv \partial^2\Omega_0/\partial\Phi^2\sim v^2\,(\Phi/v)^{\gamma-2}$. For a quartic potential, $\gamma=4$ and $|\Phi|$ increases linearly with $|\mu|$. In general, $|\Phi| \sim |\mu/v|^{2/(\gamma-2)}$ and the only way for $\Phi$ to saturate is if $m_\Phi$ abruptly increases for some reason, with $\gamma\gg 2$.

At this point, we follow \cite{Bedaque:2014sqa} and seek inspiration from QCD at large isospin chemical potential \cite{Son:2000xc,Carignano:2016lxe}, which can be simulated on the lattice \cite{Alford:1998sd,Kogut:2002zg,Brandt:2017oyy,deForcrand:2007uz} and  bears close relation to two-color QCD at finite density  \cite{Hands:2001ee,Kogut:2001na,Ratti:2004ra,Hands:2006ve,Hands:2010gd,Boz:2019enj,Astrakhantsev:2020tdl}. In that theory, the isospin-neutral chiral condensate, $\bar\sigma\sim \langle \overline{\psi}\psi\rangle$, is continuously rotated into an isospin-charged pion condensate $\bar\pi \sim \langle\overline{\psi} \gamma_5 \tau_3 \psi \rangle$ at increasing isospin chemical potential. Once the chiral condensate is completely rotated into pions, the pion condensate saturates, giving rise to a peak in the speed of sound \cite{Carignano:2016lxe}. This points us to a minimal mechanism that can lead to a speed of sound  that goes above  the conformal limit: we must introduce at least one extra field, in addition to $\Phi$. This field, which we denote by $\sigma$, must condense at lower densities and then be continuously converted into the baryon-number-charged condensate, as $\mu$ increases, up to the point where it saturates. 

Let us now devise a generic Ginzburg-Landau theory that realizes such a mechanism. 
With increasing $\mu$, the order parameter $\bm \varphi\equiv(\sigma,\Phi)$ will move along the minimum of the effective potential $\Omega(\bm\varphi,\mu)$, with $\partial\Omega/\partial\bm\varphi =0$. Thus, the total derivative
\begin{equation}
    \frac{d\;}{d\mu}\frac{\partial \Omega}{\partial \varphi_i} = \frac{\partial^2\Omega}{\partial\mu\partial\varphi_i} + \frac{\partial^2\Omega}{\partial\varphi_j\partial\varphi_i}\frac{d\varphi_j}{d\mu} =0
    \label{eq:totalmuderivative2}
\end{equation}
 must vanish. The chemical potential couples to the baryon-number-carrying component $\Phi$ according to 
\begin{equation}
    \Omega(\bm\varphi,\mu) = \Omega_0(\bm\varphi,\mu) - \mu^2\,\bm\varphi^{T}\bm Q^2 \bm\varphi/2\,,
    \label{eq:genericmucoupling}
\end{equation}
where we defined the charge matrix $\bm Q^2 \equiv \diag(0,q^2)$. 
Also, after defining the mass matrix $\bm M^2 \equiv \partial \Omega_0/\partial \bm \varphi^2$, we find
${\partial^2\Omega}/{\partial\bm\varphi^2}=\bm M^2 - \mu^2\,\bm Q^2$ and, 
from Eqs.~\eqref{eq:totalmuderivative2} and \eqref{eq:genericmucoupling}, 
\begin{equation}
    \frac{d\bm\varphi}{d\mu} = 2\,\mu\,\left(\bm M^2 - \mu^2\bm Q^2\right)^{-1}\bm Q^2\bm \varphi\,.
    \label{eq:pathwithmu}
\end{equation}

For the order parameter $\bm\varphi(\mu)$ to saturate after rotating from the $\sigma$ to the $\Phi$ direction,  $\Omega_0(\bm\varphi,\mu)$ must act as a ``gutter'', guiding $\bm \varphi$ along the path of least energy as $\mu$ increases. That is, $\bm \varphi(\mu)$ will tend to change along the direction corresponding to the smallest eigenvalue $m_<^2$ of the mass matrix, as long as it is much smaller than the largest eigenvalue $m_>^2$.  The most general (real) mass matrix with eigenvalues $m_<^2$ and $m_>^2$ can be found by applying a rotation to $\diag(m_>^2,m_<^2)$:
\begin{equation}
    \bm M^2 = \left(
\begin{array}{cc}
   m_>^2\,\cos^2\phi  + m_<^2\sin^2\phi & (m_>^2-m_<^2)\,\cos\phi\sin\phi    \\
    (m_>^2-m_<^2)\,\cos\phi\sin\phi  &  m_<^2\,\cos^2\phi  + m_>^2\sin^2\phi
\end{array}
\right)\,,
\label{eq:massmatrix}
\end{equation}
where $m_>^2$, $m_<^2$, and $\phi$ can be functions of $\bm\varphi$ and $\mu$. 
From Eqs.~\eqref{eq:pathwithmu} and \eqref{eq:massmatrix}, for $m_>^2\gg m_<^2,\mu^2$, one obtains
\begin{equation}
    \frac{d\Phi}{d\mu} \approx \frac{2\,q^2\,\mu\,\Phi\,\cos^2\phi}{ m_<^2- q^2\,\mu^2\,\cos^2\phi} +\mathcal{O}(\mu^2/m_>^2,m_<^2/m_>^2)\,.
\end{equation}
We find that $\Phi(\mu)$ saturates as soon as $\phi\to \pi/2$. The reason why becomes clear once $\phi$ is recognized as the angle between the $\sigma$ direction and the mass eigenstate of mass $m_>$. While $\cos\phi>0$, the baryon-number-charged component $\Phi$ is partially aligned with the lighter direction, of mass $m_<$, and is free to grow with increasing $\mu$. Once $\cos\phi\approx 0$, however, it becomes completely aligned with the heavy direction and stalls, with  $d\Phi/d\mu\approx 0$.

Because $\Phi$ is a bosonic field with nonzero baryon number, it is natural to interpret it as a dibaryon or as a diquark ``$\psi \psi$''. The condensate $\sigma$, on the other hand, is neutral and can be interpreted as a chiral condensate ``$\bar\psi\psi$''. In fact, an interplay such as the one we propose is verified between a $\langle q q \rangle$ and a $\langle \bar q q\rangle$ condensate in the model of Ref.~\cite{Berges:1998rc}. As a result, chiral-symmetry restauration is followed by the onset of a slowly-varying diquark gap, accomplishing the requirements of our mechanism. This resembles the case of QCD with two colors and two flavors  \cite{Hands:2001ee,Kogut:2001na,Ratti:2004ra,Hands:2006ve,Hands:2010gd,Boz:2019enj,Astrakhantsev:2020tdl}. 
One important difference is that the phase transition is  of first order in Ref.~\cite{Berges:1998rc}, while it is of second order for two-color QCD. 
A nonmonotonic speed of sound should be present in both cases, as verified already in the latter \cite{Hands:2006ve}.

\section{Gell-Mann-Majorana Symmetry}
\label{sec:symmetry}

The mechanism described above requires that a neutral condensate rotates into a baryon-number-charged condensate. The simplest way to find couplings which allow for such rotation is to conjecture an approximate symmetry between the two fields, identifying the light directions in the effective potential with pseudo-Goldstone modes. 
Since these fields have different baryon numbers, this symmetry must connect particles and antiparticles. 
In this context, it is quite natural to consider continuous transformations of the form
\begin{equation}
\Lambda_C:\quad \psi \to \psi' = \psi + i\,\epsilon\,\gamma_5\,
\psi^C
 \label{eq:transformationC}
\end{equation}
where $\epsilon$ is an infinitesimal parameter 
and $\psi^C=C\,\overline{\psi}^T$ is the charge conjugate of $\psi$ ($C=i\,\gamma^2$ in the Dirac basis of the $\gamma$ matrices).

It proves convenient to write transformation \eqref{eq:transformationC} in the Nambu-Gorkov basis $\Psi \equiv \frac{1}{\sqrt 2}(\psi,\psi^C)$:
\begin{equation}
\Lambda_C:\quad \Psi \to \Psi'=\Psi + i\,\epsilon\,G_C\,\Psi\,,\quad G_C\equiv \gamma_5\,\tau_x  \,,
\end{equation}
where we employ Pauli matrices $\vec\tau$ in the Nambu-Gorkov basis. 
Under $\Lambda_C$, Dirac bilinears transform as 
\begin{equation}
 \overline{\Psi}\,\mathcal X \Psi \to \overline{\Psi}\,\mathcal X \Psi +  i\,\epsilon\,\overline{\Psi}\,\{\mathcal X,\,G_C\} \Psi\,.
\end{equation}
As a particular case, the kinetic term in the Dirac Lagrangian $\overline{\Psi}\diag(i\slashed{\partial},i\slashed{\partial})\Psi$ is left unchanged. This is not the case for the neutral, scalar Dirac mass term
\begin{equation}
\overline{\psi}\,\psi \to \overline{\psi}\psi + i\,\epsilon\,\left(\overline{\psi}\gamma_5\,\psi^C 
+\overline{\psi}^C\gamma_5\,\psi\right),
\end{equation}
which is continuously rotated into a Majorana mass term with the quantum numbers of a dibaryon. 
Obviously, since it connects baryon and antibaryon, the transformation $\Lambda_C$ does not commute with the familiar $U(1)_B$ transformation
\begin{equation}
\Lambda_B:\quad \Psi \to \Psi'=\Psi + i\,\epsilon\,G_B\,\Psi\,,\quad G_B\equiv \tau_z\,.
\label{eq:transformationB}
\end{equation} 
Instead, we find that applying $\Lambda_B$ and $\Lambda_C$ consecutively yields a different transformation:
\begin{equation}
 \Lambda_{C^*}:\quad \Psi \to \Psi'=\Psi + i\,\epsilon\,G_{C^*}\,\Psi\,,\quad G_{C^*}\equiv\gamma_5\,\tau_y\,.
\end{equation}

Together, transformations $\Lambda_B$, $\Lambda_C$, and $\Lambda_{C^*}$ form a $SU(2)$ group, which we denote as $SU(2)_\text{GMM}$. 
While only $\Lambda_B$ is expected to be a true symmetry for charged and massive fermions, there might be situations in which an approximate realization of the full $SU(2)_\text{GMM}$ symmetry becomes relevant. This is expected to be the case for effective theories where a fraction of the Dirac mass term is generated dynamically and the physical states are not directly coupled to gauge fields. 
In such theories, the dynamically generated mass can be rotated into a superfluid gap at sufficiently high density, 
as we will see below.
This feature is largely robust against the explicit breaking of the symmetry, which makes it interesting for phenomenological applications. 

The generators $\Lambda_C$ and $\Lambda_{C^*}$ connect the chiral condensate $\sigma$, which has the quantum numbers of $\overline{\psi}\psi$, to the real and imaginary parts of a dibaryon condensate $\Phi$, which has the quantum numbers of  $\overline{\psi}\gamma_5\,\psi^C$. In turn, the generator $\Lambda_B$ rotates $\Phi$ into its complex conjugate $\Phi^*$. 
In what follows $\Phi$ will play the role of the saturating baryon-number-breaking field that generates the nonmonotonic behavior of the speed of sound, with $\sigma$ playing the role of the auxiliary condensate responsible for the saturation mechanism.

\section{Low-Energy Effective Model}
\label{sec:model}

Inspired by chiral effective models of nuclear physics, we now build an effective model in which an approximate $SU(2)_\text{GMM}$ symmetry is spontaneously broken to $U(1)_B$ at low energies. 
The pseudo-Goldstone bosons correspond to the generators  $\Lambda_C$ and $\Lambda_{C^*}$, which keep  $\sigma^2+|\Phi|^2$ fixed. These will, of course, play the role of the light excitations, of mass $m_<$, discussed in Sec.~II.  On the other hand, because of the approximate symmetry, it is natural to \emph{assume} that the Higgs-like mode---corresponding to changes in the magnitude of the condensate $(\sigma,\Phi)$--- is significantly heavier, by a factor $m_>/m_<\gg 1$. 

The low-energy degrees of freedom of the theory are the fermions and two pseudo-Goldstone bosons corresponding to the broken symmetry generators $\Lambda_C$ and $\Lambda_{C^*}$. 
By restricting ourselves to the low-energy regime, we can integrate out the heavier Higgs-like excitation, as in the nonlinear sigma model,  leaving us with the constraint $\sigma^2+|\Phi|^2= F^2$. This considerably simplifies our model, which can be written in terms of the field
\begin{equation}
 \Sigma \equiv F\,e^{i\,\gamma_5\,\vec \phi \cdot \vec \tau}= \sigma \,\mathds{1}_8 + i\,\gamma_5\,\vec\Phi\cdot\vec \tau \,,
\end{equation}
where $\vec\phi = \phi \,\hat\phi=\phi\,(\cos\alpha,\sin\alpha,0)$, $\sigma = F\,\cos\phi$, and
\begin{equation}
    \vec\Phi \equiv (\operatorname{Re}\ \Phi, \operatorname{Im}\ \Phi,0) = F\,\sin(\phi)\, \hat \phi\approx F\phi\,\hat\phi\,.
\end{equation}

The simplest fermion-meson coupling dictated by the symmetry is of the form $\overline{\Psi} \,g\,\Sigma\,\Psi$, where $g$ is the coupling constant. It can be transformed into a Dirac mass term by a suitable redefinition of the fermion field, $\Psi \to \Sigma^{1/2}\Psi/F$. This leads to the Lagrangian
\begin{equation}
 \mathcal{L} = \overline{\Psi} \,\left(i\,\gamma^\mu\partial_\mu
+ \gamma^\mu\, V_\mu
+ i\,\gamma^\mu\,\gamma_5\, A_\mu - g\,F\right)\,\Psi + 
\dfrac{F^2}{4}\tr \left[\partial^\mu U^\dagger \partial_\mu U\right]+ \dfrac{F^2\,m_\Phi^2}{4}\tr\left(U^\dagger + U\right)\,,
\label{eq:nonlinlag}
\end{equation}
where we employ $U\equiv  e^{i\,\vec \phi \cdot \vec \tau/F}$ instead of $\Sigma$ and include an explicit symmetry breaking term, yielding a mass $m_\Phi$ for the dibaryon field. Due to the redefinition of the fermion field, only derivative couplings with $\Sigma$ are kept, in the form
\begin{equation}
V_\mu = \frac{i}{2}\left(  \xi\,\partial_\mu\xi^\dagger+\xi^\dagger\,\partial_\mu\xi\right) \,,\quad
 A_\mu =\dfrac{1}{2}\left(\xi\,\partial_\mu\xi^\dagger_\sigma-\xi^\dagger\,\partial_\mu\xi\right)\,,
 \label{eq:gaugelikefields}
\end{equation}
where $\xi \equiv U^{1/2} =e^{i\,\vec \phi \cdot \vec\tau/2}$ and, in contrast with the usual nonlinear sigma model, both $A_\mu$ and $V_\mu$ are vector fields. 

The coupling to the $U(1)_B$ chemical potential $\mu$ can be found by employing the covariant derivative 
\begin{equation}
 D_\mu \Psi \equiv \partial_\mu\Psi - i\,\mu\,\delta_\mu^0\tau_z\,\Psi\,,\quad
  D_\mu U \equiv \partial_\mu U - i\, \mu\,\delta_\mu^0\,[\tau_z,U]\,.
\end{equation} 
In mean-field approximation, because the $U(1)_B$ symmetry is preserved, $\langle \alpha\rangle=0$ and
\begin{equation}
 \mathcal{L}_{MF} =\langle \overline{\Psi}\,\left(i\,\gamma^\mu\partial_\mu +\gamma^0\,\mu\,\cos\bar\phi\,\tau_z+ \gamma^0\,\gamma_5\,\mu\,\sin\bar\phi\,\tau_y - g\,F\right)\Psi\rangle +
 2\,\mu^2\,F^2\,\sin^2\bar\phi + {F^2\,m_\Phi^2}\,\cos\bar\phi\,,
 \label{eq:lagmeanfield}
\end{equation}
where we denote $\langle\phi\rangle=\bar\phi$. 
As the condensate rotates from $\bar\phi =0$ to $\bar \phi = \pi/2$, the coupling of the fermions to the chemical potential changes in form. This is because of the redefinition of the fermion field, which becomes a mixture of particle and antiparticle for $\sin\bar\phi\neq 0$. 

\begin{figure}
    \centering
    \includegraphics[width=0.4\textwidth]{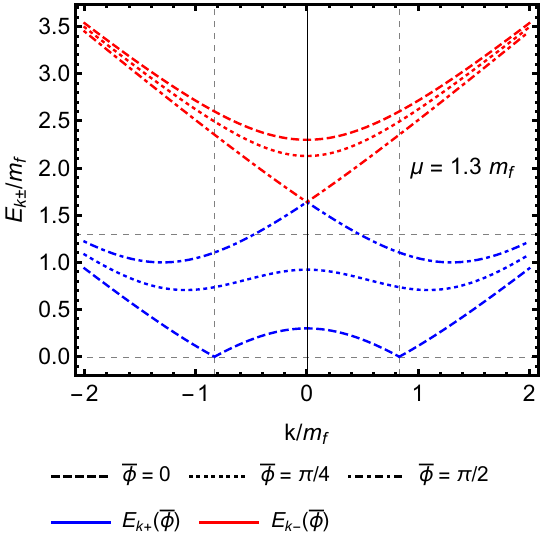}
    \caption{Dispersion relation for the fermions in the trivial ($\bar\phi=0$, dashed lines), BCS  ($\bar\phi=\pi/4$, dotted lines) and massless BCS  ($\bar\phi=\pi/2$, dash-dotted lines) phases, for $\mu = 1.3\,m_f$. The bottom  three curves correspond to particle solutions, $E_{k+}$ (in blue), while the top three ones correspond to antiparticle solutions, $E_{k-}$ (in red). See Eq.~\eqref{eq:dispersion} for the definition of $E_{k\pm}$.  }
    \label{fig:dispersion}
\end{figure}

From here on, we focus on low temperatures and approximate $T\simeq 0$. 
By evaluating the fermionic determinant, we find the effective potential at zero temperature:
\begin{equation}
\Omega(\bar\phi) = -\int \dfrac{d^3k}{(2\pi)^3}\,
 \left(E_{\bm k+}+E_{\bm k-}\right) 
 - 2\,\mu^2\,F^2\,\sin^2\bar\phi + {F^2\,m_\Phi^2}\,(1-\cos\bar\phi)\,,
 \label{eq:effpot}
\end{equation}
with
\begin{equation}
 E_{\bm k\pm}(\bar\phi) \equiv \sqrt{(\omega_{\bm k}(\bar\phi) \mp \mu)^2 + (g\,F)^2\, \sin^2\bar\phi} \,,
 \label{eq:dispersion}
\end{equation}
where $\omega_{\bm k}(\bar\phi) \equiv \sqrt{k^2+(g\,F)^2\,\cos^2\bar\phi}$. 
Equation~\eqref{eq:dispersion} interpolates continuously between the dispersion relation of free particles of mass $m_f=g\,F$ (for $\bar\phi=0$) and that of excitations in the presence of a superfluid gap $\Delta_f = g\,F$ (for $\bar\phi = \pi/2$). This dispersion relation is shown for different values of $\bar \phi$ in Fig.~\ref{fig:dispersion}.

There are two divergent contributions to the effective potential in Eq.~\eqref{eq:effpot}. The first one equals the  vacuum energy at $\mu=0$: 
\begin{equation}
    \delta \Omega_\Lambda =  -\frac{1}{\pi^2}\int_0^\Lambda k^2 dk\,\omega_{\bm k}(0)\,.
\end{equation}
and can be subtracted by shifting $\Omega$ by a constant. 
The second one is logarithmic and it reads
\begin{equation}
\delta \Omega_{\delta F^2}(\bar\phi) =
-\frac{1}{2\pi^2}(g\,F)^2\,\mu^2\,\sin^2\bar\phi\int_0^{\Lambda} \frac{k^2 dk}{\omega_{\bm k}^3(0)} \,. 
\end{equation}
This can be absorbed by renormalizing $F^2$ and $m_\Phi^2$, with $F^2\,m_\Phi^2$ kept fixed. 

A gap equation can be found by minimizing the renormalized effective potential $\Omega_R\equiv\Omega-\delta \Omega_\Lambda-\delta \Omega_{\delta F^2}$ with respect to $\bar\phi$. This yields the gap equation
\begin{equation}
   \dfrac{\partial\Omega}{\partial \bar \phi}= - 4\, \mu^2\,F^2\sin\bar\phi\,\cos\bar\phi
  +F^2\,m_\Phi^2\,\sin\bar\phi - \mu\,(g F)^2\,\sin\bar\phi\,\cos\bar\phi\,I_{\bar\phi}=0\,,
  \label{eq:condphi}
\end{equation}
which has two solutions $\bar\phi=\bar\phi^*$, with:
\begin{equation}
\left\{
 \begin{array}{l}
  \sin\bar\phi^* =0 \,, \; \textrm{ or}\\
  \cos\bar\phi^* = \dfrac{1}{\mu} \dfrac{m_\Phi^2}{4\,\mu + g^2\,I_{\bar\phi}(\bar\phi^*)}
 \end{array}\right.\,,
 \label{eq:solgapeq}
\end{equation}
where 
\begin{equation}
 I_{\bar\phi}(\bar\phi)\equiv \,\int \dfrac{d^3k}{(2\pi)^3}\dfrac{1}{\omega_k(\bar\phi)}\left(\dfrac{1}{E_{\bm k +}(\bar\phi)}-\dfrac{1}{E_{\bm k -}(\bar\phi)}-2 \,\mu \,\left(\frac{g\,F}{\omega_{\bm k}(0)}\right)^2\right)\,.
 \label{eq:Itheta}
\end{equation}
The second solution in Eq.~\eqref{eq:solgapeq} corresponds to the superfluid phase and must be found self-consistently. We note that this solution only exists for $4\,\mu^2 + g^2\,I_{\bar\phi}\geq m_\Phi^2$, which reduces to $2|\mu|\geq m_\Phi$ in the noninteracting limit $g\to 0$.

\section{Results and Discussion}
\label{sec:results}

\begin{figure*}
    \centering
    \includegraphics[width=0.9\textwidth]{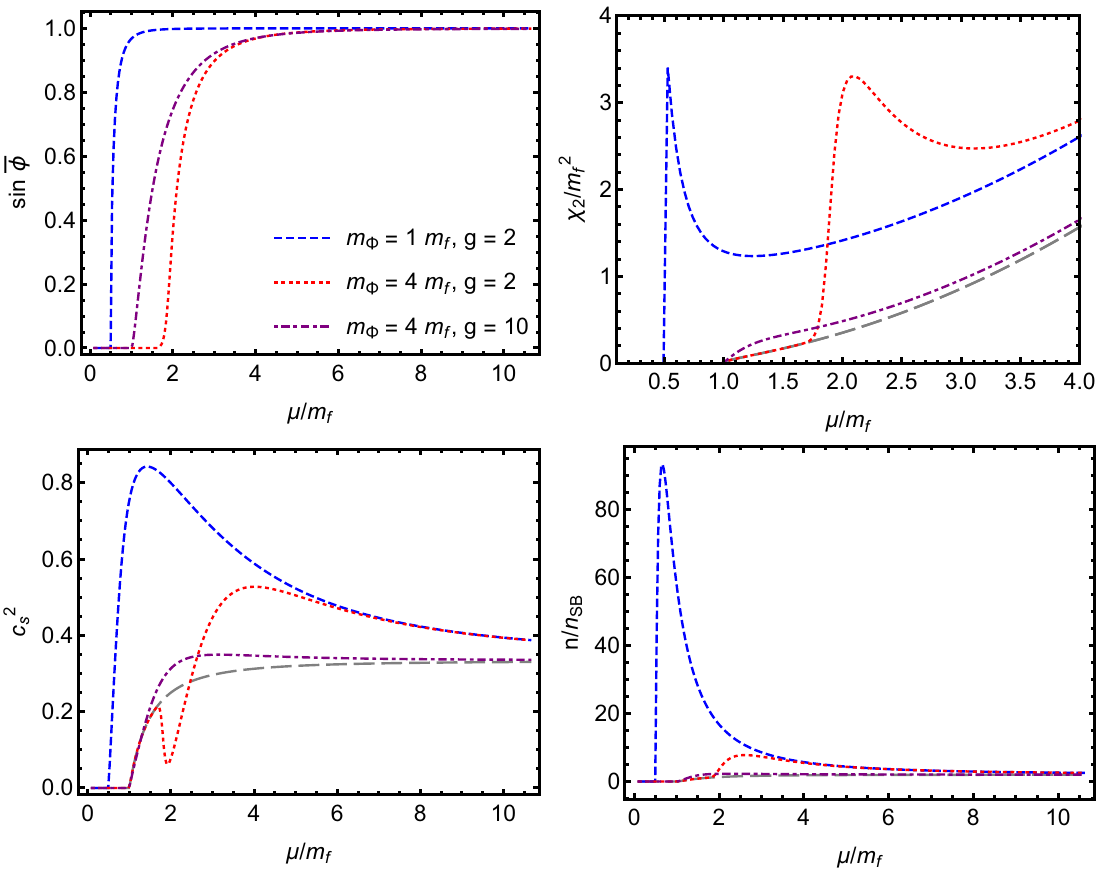}
    \caption{The order parameter $\sin \bar \phi$ (top left), the susceptibility $\chi_2$ (top right), the speed of sound $c_s$ (bottom left) and the effective number of degrees of freedom $6\pi^2\,n/\mu^3$ (bottom right), as functions of the chemical potential, normalized by $m_f=g\,F$. Dashed gray lines show results for a free gas of fermions at zero temperature, while colored lines show results for different values of the dibaryon mass $m_\Phi$ and coupling constant $g$.}
    \label{fig:results}
\end{figure*}

Having found the minimum of the effective potential, we can find the pressure $p=-\Omega_R(\bar\phi^*)$, the baryon density $n = \partial p/\partial \mu$, the energy density $\epsilon = \mu\,n - p$ and all equilibrium quantities.   
Numerical results are shown in Fig.~\ref{fig:results}. As anticipated, as $\mu$ increases, the condensate continuously rotates from the neutral to the baryon-number-charged direction. Once this rotation is finished, the expectation value $\langle |\Phi|^2 \rangle = F^2$ provides a constant contribution $\sim 4 F^2$ to the susceptibility, which in turn leads to a peak in the speed of sound, which goes above the conformal limit. The peak in the speed of sound can also be explained in terms of the effective number of degrees of freedom $n/n_\text{SB}=6\pi^2\,n/\mu^3$, which continuously decreases toward the two spin states of the fermion field once the baryon-number-carrying condensate has saturated \cite{Bedaque:2014sqa}. 

The contribution of the condensate to the susceptibility can be calculated by replacing the second solution in Eq.~\eqref{eq:solgapeq} into the second derivative of the effective potential:
\begin{equation}
 \chi_{2}^\text{cond} =   12 \,F^2\,\left(\frac{m_\Phi}{2\mu}\right)^4+4\,F^2\;\;\; (+ \text{fermionic contribution}) \,.
 \label{eq:chi2cond}
\end{equation}
Right above the critical chemical potential, $\chi_2$ peaks, causing a dip in the speed of sound. Then, as the condensate rotates toward the baryon-number-charged direction, the first term in Eq.~\eqref{eq:chi2cond} dominates and $\chi_2$ decreases, restoring the speed of sound to its previous value. Finally, after the dibaryon condensate saturates, the second term in Eq.~\eqref{eq:chi2cond} dominates, leading to a  flat (or minimum) susceptibility and a peak in the speed of sound. At even higher values of $\mu$, the fermionic contribution dominates and the speed of sound decreases toward the conformal limit from above.

The arguments above depend on a few steps of decreasing generality. To draw appropriate conclusions, we must clarify the importance of each assumption and choice. That a large, nonperturbative energy scale must contribute to the baryon-number susceptibility to produce a large peak in the speed of sound seems inescapable. Only in this way it is possible to circumvent arguments from dimensional-analysis. If we consider the same arguments from the point of view of the number of spatial dimensions, with $n\sim \mu^d$, it is natural to relate this energy scale to a lower number of effective dimensions on the Fermi surface---say, due to a gap hindering its growth. In quarkyonic matter, for instance, this scale is provided by the depth of the baryonic shell, $\Delta \sim \Lambda_\text{QCD}$, on top of the quark Fermi surface \cite{McLerran:2007qj,McLerran:2018hbz,Jeong:2019lhv,Zhao:2020dvu,Margueron:2021dtx,Duarte:2021tsx}. When the Fermi momentum of the quarks $k_F\sim \mu \ll \Delta$, the leading contribution to the density becomes $n\sim \Delta^2\,k_F\sim \Delta^2\,\mu$  \cite{McLerran:2018hbz,Zhao:2020dvu,Duarte:2021tsx}. 
Here, instead of a gap on top of the quark Fermi surface due to the formation of baryons, we consider a gap on top of the the baryonic Fermi surface due to an instability against the formation of a dibaryon condensate. 
We view this approach as more conservative, since it does not require quark degrees of freedom. Nevertheless, the nature of the degrees of freedom is not essential to the mechanism and a gap due to the formation of diquarks would do just as well. Implications of diquark condensation to the speed of sound in dense matter were made in \cite{Leonhardt:2019fua,Roupas:2020nua}.

Once we assume the nonperturbative energy scale to emerge from a diquark or dibaryon gap, a mechanism for saturating  this gap is also required. Thus, another robust point in our argument corresponds to the need of an interplay between heavy and light degrees of freedom. On the other hand, it is not necessary that these degrees of freedom display any sort of symmetry. Other mechanisms could be responsible for saturating a baryon-number-carrying condensate, related or not to symmetry groups that break baryon-number conservation. At any rate, and even though it is not essential to our reasoning, the Gell-Mann--Majorana symmetry of Sec.~\ref{sec:symmetry} is appealingly simple. 
It also bears a clear resemblance to the chiral symmetry found in two-color, two-flavor QCD \cite{Hands:2001ee,Kogut:2001na,Ratti:2004ra,Hands:2006ve,Hands:2010gd,Boz:2019enj,Astrakhantsev:2020tdl}. A number of extensions to the $SU(2)_\text{GMM}$ group can be found by combining flavor, chiral and charge-conjugation transformations, which could be of interest in phenomenological applications.

Bearing in mind all of the above observations, it seems very plausible that a peak in the speed of sound of cold and dense matter is caused by the condensation of baryon or quark Cooper pairs, catalyzed by the presence of a neutral condensate. This presents an alternative to the mechanism realized in quarkyonic matter \cite{McLerran:2007qj,McLerran:2018hbz,Jeong:2019lhv,Zhao:2020dvu,Margueron:2021dtx,Duarte:2021tsx}, with no need for quarks or any exotic degrees of freedom. It could also provide a simple explanation for the speed-of-sound peak found in Refs.~\cite{Kojo:2014rca,Leonhardt:2019fua,Baym:2019iky,Roupas:2020nua,Malfatti:2020onm,Ayriyan:2021prr}. 
Because of the spontaneous breakdown of the $U(1)_B$ symmetry at large $\mu$, this mechanism is expected to display a phase transition. However, the baryon-number susceptibility does not diverge, nor does the speed of sound vanish  in our model. Even so, for a large range of parameters, the susceptibility still peaks, leading to a dip before the peak in the speed of sound (see Fig.~\ref{fig:results}). This seems to be a unique feature of our mechanism. If present in the EoS of cold and dense matter, this speed-of-sound dip could lead to a softer EoS at slightly lower densities.

The lack of divergences in $\partial^2 p/\partial \mu^2$ can be explained by noting that the chemical potential itself does not break the $U(1)_B$ symmetry explicitly. If an external field $H$ were included so as to explicitly break $U(1)_B$, the corresponding susceptibility $\partial^2 p/\partial H^2|_{H=0}$ would diverge at the onset of the $U(1)_B$-breaking condensate. It could prove instructive to analyze the behavior of the specific heat across the transition. We note that, very close to the transition, large fluctuations are expected and the mean-field approximation becomes unreliable, so that we cannot establish its nature. However, we note that the mean-field treatment and the specifics of the toy model of Sec.~\ref{sec:model} are not essential to the main physical mechanism put forth in Sec.~\ref{sec:generalarguments}.

\section{Conclusions}
\label{sec:conclusions}

In this paper, we have addressed possible ramifications of the conjectured peak in the speed of sound of cold and dense matter. First, we have identified a necessary ingredient for the emergence of such feature in the EoS of dense matter: the onset of a large, nearly constant, nonperturbative energy scale contributing to the baryon-number susceptibility. Furthermore, we have identified a class of models capable of attaining this condition. We argue that such models would arise naturally in the interface between a phase presenting a large chiral condensate and one featuring a superfluid or superconducting gap, which we dub the ``chiral-superfluid'' transition. The relation between chiral symmetry and superconducting gaps at finite density were previously explored, for instance, in  \cite{Alford:1997zt,Berges:1998rc,Hosek:1998un,Hands:2001ee,Kogut:2001na,Ratti:2004ra,Blaschke:2007ri,Fukushima:2021ctq}. The mechanism put forth here can be related to the one found in quarkyonic matter, where the role of the gap would be played by the depth of the baryonic shell of the Fermi sphere \cite{McLerran:2007qj,McLerran:2018hbz,Jeong:2019lhv,Zhao:2020dvu,Margueron:2021dtx,Duarte:2021tsx}. While quarkyonic matter is driven by large-$N_c$ physics, one could argue that our proposal is inspired by the physics of $N_c=N_f=2$ QCD  \cite{Hands:2001ee,Kogut:2001na,Ratti:2004ra,Hands:2010gd,Boz:2019enj,Astrakhantsev:2020tdl}, where baryons are diquarks \cite{Hands:2010gd}. 

We then turned to an explicit model implementation of this scenario. For greater simplicity, we introduced an auxiliary symmetry between the chiral condensate and the superfluid gap, consisting of a $SU(2)_\text{GMM}$ group. By imposing this symmetry to be approximately realized, we arrived at a minimalistic low-energy effective model of the chiral-superfluid transition. As expected, we found that the EoS of this model presents a clear peak in the speed of sound, capable of reaching values above $c_s^2=1/3$. For some sets of parameters, we found that this peak is preceded by a speed-of-sound dip, similar to the one found in the QCD high-temperature crossover \cite{Borsanyi:2013bia,Bazavov:2014pvz}. The structure in the speed of sound corresponds to a minimum in the baryon-number susceptibility, sometimes following a peak, which we attribute to a large constant contribution to the susceptibility. In different model realizations, diquarks might condense in a first-order or second-order transition, leading to a softer EoS before the peak in the speed of sound. We note that, in our specific implementation, the conformal limit $c_s^2\to 1/3$ is achieved from above and not from below. We attribute this to the incomplete and simplified nature of the model, which could be improved, say, by including other interactions. We stress that the toy model of Sec.~\ref{sec:model} is not intended to be realistic. We leave the task of constructing a more realistic implementation of the same mechanism, which would be suitable for investigations of the neutron star mass-radius relation, for future work.  

We do not claim that the developments hereby presented are the only possible solution to the conundrum set by recent neutron-star observations. Rather, we view it as a simple, conservative  alternative to other proposals \cite{McLerran:2007qj,Drago:2013fsa,McLerran:2018hbz,Zhao:2020dvu,Pisarski:2021aoz,Margueron:2021dtx,Duarte:2021tsx,Bombaci:2020vgw,Traversi:2021fad,Annala:2019puf}, in the sense that no exotic degrees of freedom are necessary. In particular, our mechanism naturally highlights the role played by superfluid or superconducting gaps in ultradense matter, and it might help explain the physics responsible for the emergence of a nonmonotonic speed of sound in  previous model calculations \cite{Kojo:2014rca,Leonhardt:2019fua,Baym:2019iky,Roupas:2020nua,Malfatti:2020onm,Ayriyan:2021prr}. While it is well known that diquarks can make the EOS of dense matter stiffer (see e.g., Ref~\cite{Kojo:2014rca}), the importance of the interplay with other degrees of freedom has not, to our knowledge, been pointed out before. This could provide valuable  guidance for further models of neutron-star matter.

\section*{Acknowledgments} 
We thank J. Noronha-Hostler for comments on the neutron star mass-radius relations. M.H. thanks D.~Kroff for fruitful discussions. 
E.S.F. is partially supported by CAPES (Finance Code 001), CNPq, FAPERJ, and INCT-FNA (Process No. 464898/2014-5).
J.N. is partially supported by the U.S. Department of Energy, Office of Science, Office for Nuclear Physics under Award No. DE-SC0021301.

\bibliography{speedofsound}
\end{document}